\documentstyle[preprint,aps,amssymb]{revtex}
\begin{document}
\title{Closed-Form Expressions for the Noncompact Part of $Sp(2n)$}
\author{Ivan Gjaja}
\address{\centerline{Theoretical Division and 
Center for Nonlinear Studies, Los Alamos National Laboratory,}
Los Alamos, NM 87545}
\maketitle
\begin{abstract}
\baselineskip 20pt
The Iwasawa decomposition of the 
symplectic group is used to write down
closed-form expressions for the noncompact part of the group. The results 
make possible the computation of Lyapunov exponents using the method 
described in \cite{habib} in systems with more than two degrees of freedom.
\end{abstract}
\vspace{.2in}

In these Proceedings \cite{habib} Habib and Ryne propose a method for
computing Lyapunov exponents in cases where the linearized dynamics is 
Hamiltonian. The method requires a closed-form expression for the noncompact 
part of the symplectic group, denoted in \cite{habib} by $Sp(2n)$, 
in the defining representation. 
(Other common notations for the symplectic group are 
$Sp(n,{\Bbb R})$ and $Sp(2 n,{\Bbb R})$.) The authors  
provide such an expression for $Sp(2)$ and $Sp(4)$ and state that 
they are working on generalizations to $Sp(2n)$. In this note I
point out that suitable closed-form expressions for the noncompact part 
of $Sp(2n),\;\;n\ge 1$, can be found rather easily using the Iwasawa 
decomposition of this group
\cite{barut}. 

Denoting the set of matrices of the form $J S_c$ and $J S_a$ 
(Eq. (10) of \cite{habib}) by $K$ and $P$, I first demonstrate that 
they represent the Cartan decomposition of the symplectic Lie algebra 
$sp(2n)$. In the calculations that
follow use is made of the Killing form for $sp(2n)$ which is given by
$(X_1,X_2)=(2 n+2){\rm Tr}(X_1 X_2)$ for any $X_{1,2}\in sp(2n)$ 
\cite{helgason}.
Then, since any element of $K$ can be written in the form $JS+SJ$
where $S$ is a symmetric $2n\times 2n$ matrix and $J$ is defined in
\cite{habib}, and any element of $P$ in the form $JS-SJ$ \cite{dragt},
it can be verified that
$[K,K]\subset K$, $[K,P]\subset P$,
$[P,P]\subset K$, $(X,X)<0$ for $X\ne 0,\;\;X\in K$, and
$(Y,Y)>0$ for $Y\ne 0,\;\;Y\in P$. Thus $sp(2n)=K+P$ is indeed the Cartan 
decomposition of $sp(2n)$. $K$ is the maximal
compact subalgebra, which in this case is the unitary Lie algebra $u(n)$. 
To compute the closed-form expression for the noncompact factor of
$Sp(2n)$ displayed in Eq. (11) of \cite{habib}, I begin by writing down the 
Iwasawa decomposition
of $sp(2n)$, $sp(2n)=u(n)+H_P+N$. $H_P$ is an abelian
and $N$ a nilpotent Lie algebra for which the generators can be taken as
\begin{mathletters}
\begin{equation}
h_i=e_{ii}-e_{n+i,n+i},\;\;\;\;1\le i\le n\;\;\;\;{\rm for}\;\; H_P;
\label{eq:1a}
\end{equation}
\begin{equation}
m_{ij}=e_{i,n+j}+e_{j,n+i},\;\;\;\;1\le i,j \le n
\label{eq:1b}
\end{equation}
and
\begin{equation}
n_{ij}=e_{ij}-e_{n+j,n+i},\;\;\;\;1\le i,j \le n,\;\;\;\;i<j\;\;\;\;
{\rm for}\;\; N.
\label{eq:1c}
\end{equation}
\end{mathletters}
Here $e_{ij}$ is the usual Weyl basis for $2n\times 2n$ matrices,
$(e_{ij})_{rs}=\delta_{ir}\delta_{js}$. (The method for computing the
Iwasawa decomposition for any semisimple real Lie algebra is given in
\cite{barut}.)

An arbitrary element of $Sp(2n)$ can now be written as
$e^X e^Z e^W$, where $X\in u(n),\; Z\in H_P,\; W\in N$ \cite{helgason}.
Since $H_P$ is abelian, the exponential series in $e^Z$ is easily evaluated, 
whereas the nilpotency of $N$ converts the infinite series in $e^W$ into a 
polynomial of order $2 n-1$. The factor $e^Z e^W$ is thus
\begin{equation}
e^Z e^W={\rm diag}(\lambda_1,\ldots,\lambda_n,
\lambda_1^{-1},\ldots,\lambda_n^{-1})\Big(I+\sum_{s=1}^{2n-1}
{1\over {s!}}\big(\sum_{i,j} a_{ij}m_{ij}+\sum_{i<j}
b_{ij} n_{ij}\big)^s\Big).
\label{eq:2}
\end{equation}
The parameters $a_{ij}$ and $b_{ij}$ are in ${\Bbb R}$ and 
$\lambda_1,\ldots\lambda_n^{-1}>0$. The quantity of interest in \cite{habib},
$M\tilde M$, takes the form
\begin{equation}
M\tilde M=\widetilde{\Big( e^Z e^W\Big)} e^Z e^W.
\label{eq:3}
\end{equation}

As an example of the use of Eq. (\ref{eq:2}) I write down the matrix display 
for $Sp(4)$:
\begin{equation}
e^Z e^W={\rm diag}(\lambda_1,\lambda_2,\lambda_1^{-1},\lambda_2^{-1})
\left(\matrix{
1&b_{12}&2 a_{11}-{1\over 3}a_{22}b_{12}^2&a_{12}+a_{22}b_{12}\cr
0&1&a_{12}-a_{22}b_{12}&2 a_{22}\cr 0&0&1&0\cr 0&0&-b_{12}&1\cr}\right).
\label{eq:4}
\end{equation}

\end{document}